\documentclass[%
 aip,
rsi,%
 amsmath,amssymb,
 a4paper,
 reprint,%
]{revtex4-1}

\usepackage{graphicx}
\usepackage{dcolumn}
\usepackage{bm}
\usepackage{float}
\usepackage[english]{babel}

\begin{document}

\preprint{ArXiv}

\title[Nonlinear filtering of an optical pulse train using dissipative Kerr solitons]{Nonlinear filtering of an optical pulse train using dissipative Kerr solitons}

\author{Victor Brasch}
\affiliation{CSEM, CH-2000, Neuchatel, Switzerland}
\author{Ewelina Obrzud}
\affiliation{CSEM, CH-2000, Neuchatel, Switzerland}
\affiliation{University of Geneva, Department of Astronomy, CH-1290, Versoix, Switzerland}
\author{Steve Lecomte}
\affiliation{CSEM, CH-2000, Neuchatel, Switzerland}
\author{Tobias Herr}
\email{tobias.herr@csem.ch}
\affiliation{CSEM, CH-2000, Neuchatel, Switzerland}

\begin{abstract}
The capability to store light for extended periods of time enables optical cavities to act as narrow-band optical filters, whose linewidth corresponds to the cavity's inverse energy storage time. Here, we report on \textit{nonlinear} filtering of an optical pulse train based on temporal dissipative Kerr solitons in microresonators. Our experimental results in combination with analytical and numerical modelling show that soliton dynamics enables storing information about the system's physical state longer than the cavity's energy storage time, thereby giving rise to a filter width that can be more than an order of magnitude below the cavity's intrinsic linewidth. Such nonlinear optical filtering can find immediate applications in optical metrology, low-timing jitter ultra-short optical pulse generation and potentially opens new avenues for microwave photonics.
\end{abstract}

\maketitle

\section{Introduction}
Cavities are omnipresent in optics. For instance, lasers rely on the feedback by their cavity \cite{maiman_stimulated_1960} or can be stabilized to a cavity \cite{drever_laser_1983}. Ultra-stable resonators have enabled sub-Hz linewidth lasers and represent the fly-wheels for optical atomic clocks \cite{ludlow_optical_2015}. In particular, filter cavities have been used to reduce the timing jitter of pulsed laser sources and optical signals, which is of high importance to microwave (MW) photonics and optical precision metrology \cite{capmany_microwave_2007, beha_electronic_2017}. One of the key characteristics of optical cavities is their linewidth $\kappa$ and the related cavity decay time $1/\kappa$. Any dynamics faster than $1/\kappa$ (i.e. Fourier-frequencies in excess of $\kappa$) are suppressed via the linear superposition and temporal averaging of the light fields inside the cavity. Beyond linear optics, the strong light intensity in optical enhancement cavities has enabled efficient nonlinear optical frequency conversion and the advent of high-Q dielectric optical microresonators \cite{vahala_optical_2003} has given rise to several new areas of research in nonlinear optics.
\begin{figure}[h]
\centering
\includegraphics[width=0.95\linewidth]{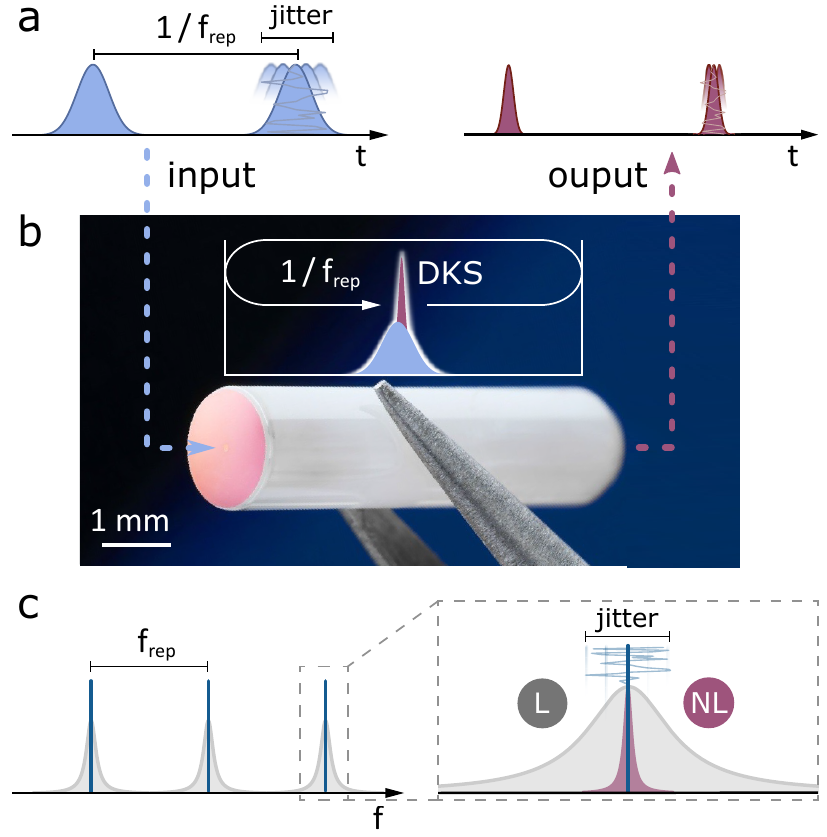}
\caption{\textbf{Nonlinear filtering of an optical pulse train. a} The Fabry-Pérot filter reduces the timing jitter of the pulse train. \textbf{b} Photograph of a fiber-based Fabry-Pérot microresonator as it was used in this work. The highly reflective dielectric mirror appears pink. Sketched into the image is a dissipative Kerr soliton (DKS) circulating inside the cavity on top of a driving pulse with a round-trip time of $1/f_\mathrm{rep}$. \textbf{c} Linear (L) and nonlinear (NL) filtering effects in the frequency domain. The linear filtering effect is given by the Lorentzian lineshape and the linewidth of the cavity (grey). The nonlinear filtering effectively provides a substantially narrower filter (purple).}
\label{fig:fig1}
\end{figure} 
One of these areas of research, which has gained a lot of attention over the last years, is the one of temporal dissipative Kerr solitons (DKS) in microresonators \cite{herr_temporal_2013} that are coherently driven by a continuous wave (CW) laser or by synchronous optical pulse trains \cite{obrzud_temporal_2017}. DKS are pulses of light, which retain their shape while propagating in the cavity as long as the driving laser is present, owing to a balance between dispersion and the focusing part of the nonlinearity as well as between the optical losses and nonlinear gain \cite{akhmediev_dissipative_2008}. Since their first experimental observation in long, fiber-based cavities \cite{ leo_temporal_2010} and crystalline microresonators \cite{herr_temporal_2013} as well as microresonators of other types \cite{brasch_photonic_2016, yi_soliton_2015}, DKS have proven to be a breakthrough in particular for the field of microresonator-based Kerr frequency combs \cite{delhaye_optical_2007, weiner_frequency_2017, kippenberg_dissipative_2018, gaeta_photonic-chip-based_2019}. Important for experimental systems, DKS have demonstrated to be stable and robust against small changes of their environment and to tolerate perturbations of the ideal system \cite{herr_mode_2014, sahoo_complete_2019}. It is moreover expected that DKS benefit from being generated inside a cavity, which filters noise that may be present in the driving laser. Interestingly, and so far unexplained, it has been observed in a weakly phase-modulated CW-driven microresonator \cite{weng_spectral_2019} that DKS can lead to a reduction of microwave modulation phase noise at noise frequencies below $\kappa$. In addition, sub-linewidth suppression of amplitude noise was observed for a synchronously-driven DKS in free-space enhancement cavities \cite{lilienfein_temporal_2019}. 

In this work, we generate DKS in fiber-based Fabry-Pérot microresonators (Fig.~\ref{fig:fig1}b) via synchronous pulsed driving. Such pulsed driving has enabled unprecedentedly high conversion efficiency, deterministic operation and, importantly, all-optical stabilization of the DKS via slow synchronization to the driving pulses with metrological precision \cite{obrzud_microphotonic_2019}. A major open question, however, is how timing jitter in the input pulse train translates into DKS timing jitter (Fig.~\ref{fig:fig1}a). Here, besides the expected \textit{linear} cavity filtering effect, we find that DKS lead to an additional drastic reduction of timing fluctuations already well within the cavity's bandwidth $\kappa$. 
Despite the differences in the investigated DKS systems, phenomenologically this observation is similar to the observations mentioned above \cite{weng_spectral_2019, lilienfein_temporal_2019}, suggesting a similar physical origin.
In order to explain the observations, we show analytically, based on soliton physics, that DKS give rise to \textit{nonlinear} optical filtering equivalent to an intrinsic second filter cavity with an effective linewidth $\kappa_\mathrm{NL}\ll\kappa$ (Fig.~\ref{fig:fig1}c).

Combining both, linear and nonlinear filtering, into a single model, we are able to understand the experimentally obtained data and explain the observed filter characteristics. Moreover, numerical simulations confirm the model and allow us to relate different measured $\kappa_\mathrm{NL}$ and features in the optical spectra to the relative temporal position of the DKS with respect to the intracavity driving pulse. As such, we establish a novel connection between the nonlinear optical physics of DKS and the use of cavities as optical filters. 
\begin{figure*}[ht!]
\centering
\includegraphics[width=1\linewidth]{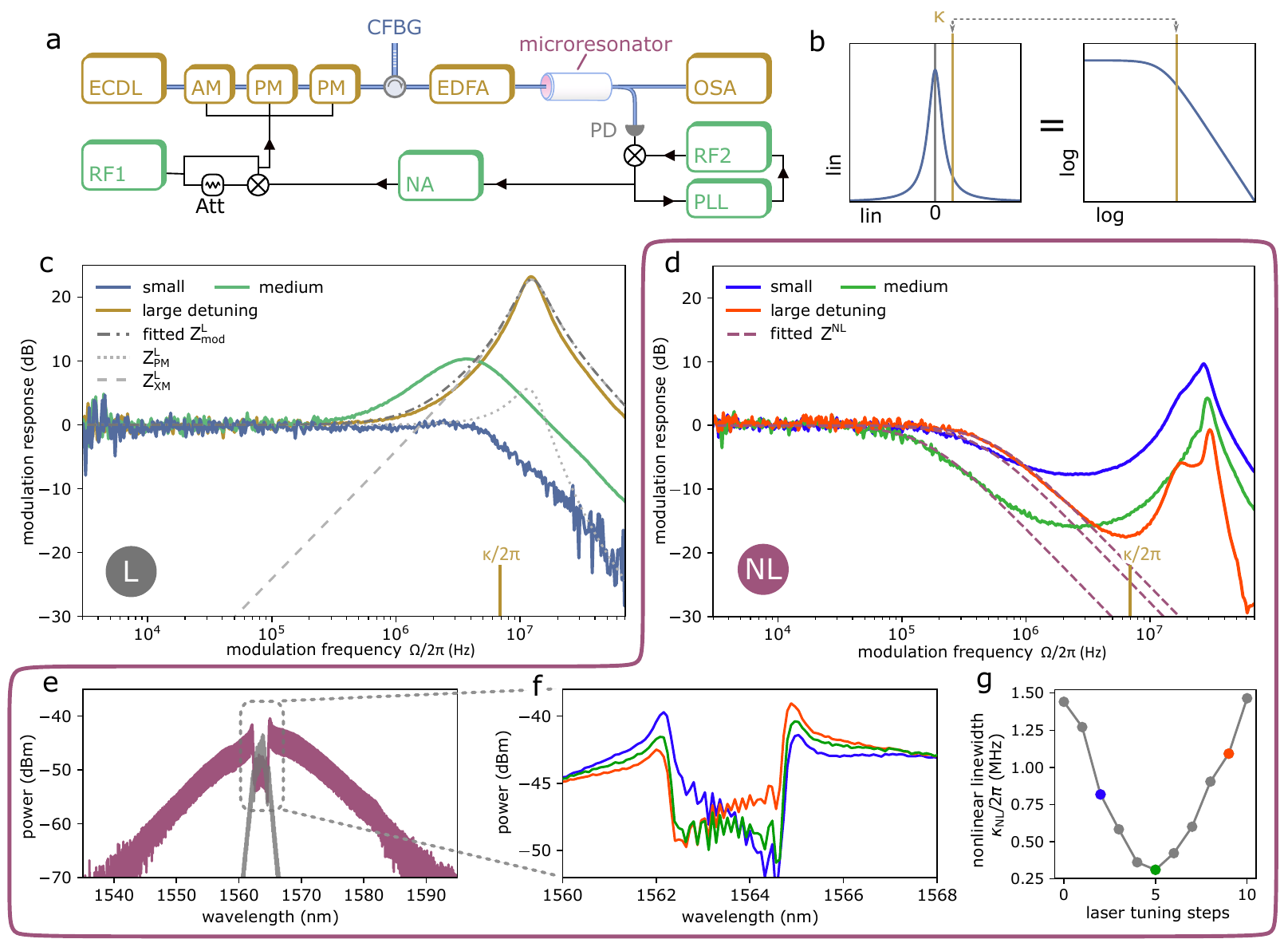}
\caption{\textbf{Nonlinear filter characterization. a}
Setup used to characterize the linear and nonlinear filtering effects. AM: electro-optic amplitude modulator; Att: RF attenuator; CFBG: chirped fiber Bragg grating; ECDL: external-cavity diode laser; EDFA: erbium-doped fiber amplifier; Microresonator: fiber-based Fabry-Pérot microresonator; NA: network analyzer; OSA: optical spectrum analyzer; PD: photodetector; PLL: phase-locked loop controller; PM: electro-optic phase modulator; RF: MW synthesizer. 
\textbf{b} Lorentzian lineshape on a linear scale and on a double-logarithmic scale. 
\textbf{c} Phase modulation (PM) response of the cavity in the linear regime for three different laser detunings. The traces are measured with the NA and are then normalized. The dashed-dotted line is the fitted full linear model $Z_\mathrm{mod}^\mathrm{L}(\Omega)$. The dotted and dashed lines show the individual contributions of $Z^\mathrm{L}_\mathrm{PM}$ and $Z^\mathrm{L}_\mathrm{XM}$ respectively as explained in the text.
\textbf{d} Nonlinear filter response with a DKS state present in the cavity for three different detunings (blue, green, red). Purple lines represent the respective best fit of $Z^\mathrm{NL}$ to the low frequency part of the modulation response. The resulting values of $\kappa_\mathrm{NL}$ are given in \textbf{g}.
\textbf{e} Optical spectra of the input (grey) and the output after the microresonator in a DKS state (purple). 
\textbf{f} Envelope of the DKS spectrum around the input spectrum for different detunings corresponding to \textbf{d}. 
\textbf{g} Fitted values for $\kappa_\mathrm{NL}$ for different detunings. The x-axis is not a quantified detuning but represents different measurements for different driving laser positions as defined by a change in the laser lock point. This monotonically changes the detuning. Colors represent the same detunings as in \textbf{d} and \textbf{f}.}
\label{fig:fig2}
\end{figure*}
\section{Experimental Setup}
\label{sec:dks}
For the experiments we use a fiber-based, anomalous-dispersion Fabry-Pérot microresonator (Fig.~\ref{fig:fig1}b) \cite{obrzud_temporal_2017} with a linewidth of $\kappa/2\pi \approx 7$~MHz and a free-spectral range (FSR) of 9.8~GHz. This microresonator is synchronously-driven by an input pulse train whose pulse repetition rate $f_\mathrm{rep}$ is approximately matched to the cavity's FSR such that a DKS is generated when the laser is tuned into resonance \cite{obrzud_temporal_2017}. Experimentally, a small repetition rate mismatch $\Delta=f_\mathrm{rep} - \mathrm{FSR}$ is usually present and, in particular, dependent on the pump power and the laser detuning $\delta$ as different intracavity power levels thermally induce a change of the FSR.

The driving pulse train is synthesized via electro-optic modulation from a continuous wave laser \cite{kobayashi_optical_1988}. To this end, the CW external-cavity diode laser (ECDL) is fed into a cascade of one electro-optic amplitude modulator and two electro-optic phase modulators (Fig.\ref{fig:fig2}a), driven by the same MW source, with suitable phase delays. A chirped fiber Bragg grating and a spool of single-mode fiber are used for pulse compression. Before coupling the light to the cavity, the pulse train is amplified to an average power of around 100 to 500~mW using an erbium-doped fiber amplifier (EDFA). Using the Fabry-Pérot microresonator in transmission allows us to readily measure the intra-cavity field, rejecting light-components that are not coupled to the cavity (corresponding to a drop port in ring-type microresonators \cite{wang_drop-port_2013}). After the cavity the signal is split and in one part detected with an optical spectrum analyzer; the other part is amplified with a second EDFA (not shown in Fig.~\ref{fig:fig2}a), detected on a fast photodetector and further amplified in the MW domain.

In our experiment, we aim at characterizing the response of the DKS to timing jitter in the input pulse train. In order to avoid contamination from noise sources in the experiment we measure the response in a coherent manner. An artificial timing modulation is introduced in the input pulse train and the response of the out-coupled pulse train is measured.
As the timing modulation in the optical domain is identical to a phase modulation in the MW domain, we practically generate the timing modulation by imprinting a phase modulation onto the MW signal driving the electro-optic modulators. This phase modulation (PM) is accomplished via a MW Mach-Zehnder interferometer where the second arm is strongly attenuated and in quadrature with the first arm. A mixer in the second arm is used to electrically vary its transmission and thereby to modulate the phase of the MW signal after the interferometer. The mixer's modulation input is connected to the output of a network analyzer (NA). The input of the NA is connected to the demodulated MW pulse repetition rate signal after the cavity. As slow thermal drifts in the optical setup impact the phase of the microwave signal after the photodetector, the demodulation is accomplished via a second reference MW synthesizer that is slowly phase-locked to the detected and amplified MW signal.

In order to derive the modulation response $Z_\mathrm{meas}(\Omega)$ at modulation frequency $\Omega$ from the NA data, the raw traces are normalized by a reference trace taken with the same optical and MW paths but without the optical cavity and with a matching photo-current on the photodetector. In addition, because the internal amplitude referencing of the NA is not meaningful in our measurements, we normalize the measured response functions to 0~dB at the near-DC end of the measurement.
\section{Experimental Results}
In a first step we characterize the linear filtering characteristics of the cavity when the input pulses are at low optical power, such that no nonlinear effects are observed. Using the transmitted power as an error signal, the pulsed driving laser is locked at different detunings. Here, we define the detuning $\delta = \omega_{0}-\omega_\mathrm{p}$ as the difference between the central spectral frequency component $\omega_\mathrm{p}$ of the driving pulses and the closest resonance frequency $\omega_{0}$ of the cavity. 

The simplest case is the one of zero detuning, $\delta = 0$, where the expected measured modulation response $ Z_\mathrm{meas}^\mathrm{L}$ is a Lorentzian-shaped filter with a cut-off frequency given by the linewidth $\kappa$ as shown in Fig.~\ref{fig:fig2}b. This is also what we observe experimentally for $\delta \approx 0$ (dark blue trace in Fig.~\ref{fig:fig2}c). For larger detunings, $ Z_\mathrm{meas}^\mathrm{L}$ does not exhibit the simple Lorentzian lineshape anymore (Fig.~\ref{fig:fig2}c, green and golden traces). Instead, the modulation response shows a peak at around the detuning frequency caused by different effects, which can be accurately described by an analytic model (dashed-dotted line) as we show below.

In a second step, we repeat the previous measurements in the nonlinear optical regime when a single DKS pulse is present in the cavity (Fig.~\ref{fig:fig2}d).

Once created, the DKS pulse is stable over a certain range of pump power levels, laser detunings $\delta$ and repetition rate mismatches $\Delta$ \cite{obrzud_temporal_2017}. In order to create reproducible experimental conditions, we adjust the repetition rate $f_\mathrm{rep}$ such that the DKS spectrum is as symmetric as possible for a medium pump laser detuning $\delta$.

Markedly different from the linear optical case, we see a substantial suppression at frequencies well below the cavity's linear linewidth $\kappa$ (golden-colored marks in Fig.~\ref{fig:fig2}c, d). This suppression can reach values of more than 10~dB for frequencies ten times lower than $\kappa$. The strength and the frequency onset of this suppression depend on the detuning $\delta$ as shown exemplary for three cases (red, green, blue) in Figure~\ref{fig:fig2}d. 

Similar to the linear case we also see a local enhancement of the modulation response for values around the detuning, albeit significantly weaker than in the linear case and with a double-peak structure, which is only resolved for certain detunings. 

Simultaneously to measuring the transfer of the timing modulation, we record the intracavity optical spectra at the output port. In the central portion of these spectra, where the DKS and the input pulse spectrum overlap (purple and grey in Fig.~\ref{fig:fig2}e respectively), we observe a striking correlation between the spectrum and the strength of the nonlinear filtering. This is shown in Figure~\ref{fig:fig2}f where the same colors in panels d and f correspond to the same DKS state with the same detuning. For the lowest frequency filtering onset (green color), the envelope of the central portion of the spectrum is almost flat whereas it is tilted towards longer or shorter wavelengths otherwise. Experimentally, the soliton state can be adiabatically moved between those states by changing the detuning or by slight changes of the input pulse repetition rate. We note that a perfectly flat central part of the spectral envelope is typically not observed and a switching behavior between long and short wavelength tilt (and vice versa) is commonly the case.
\begin{figure*}[ht!]
\centering
\includegraphics[width=1\linewidth]{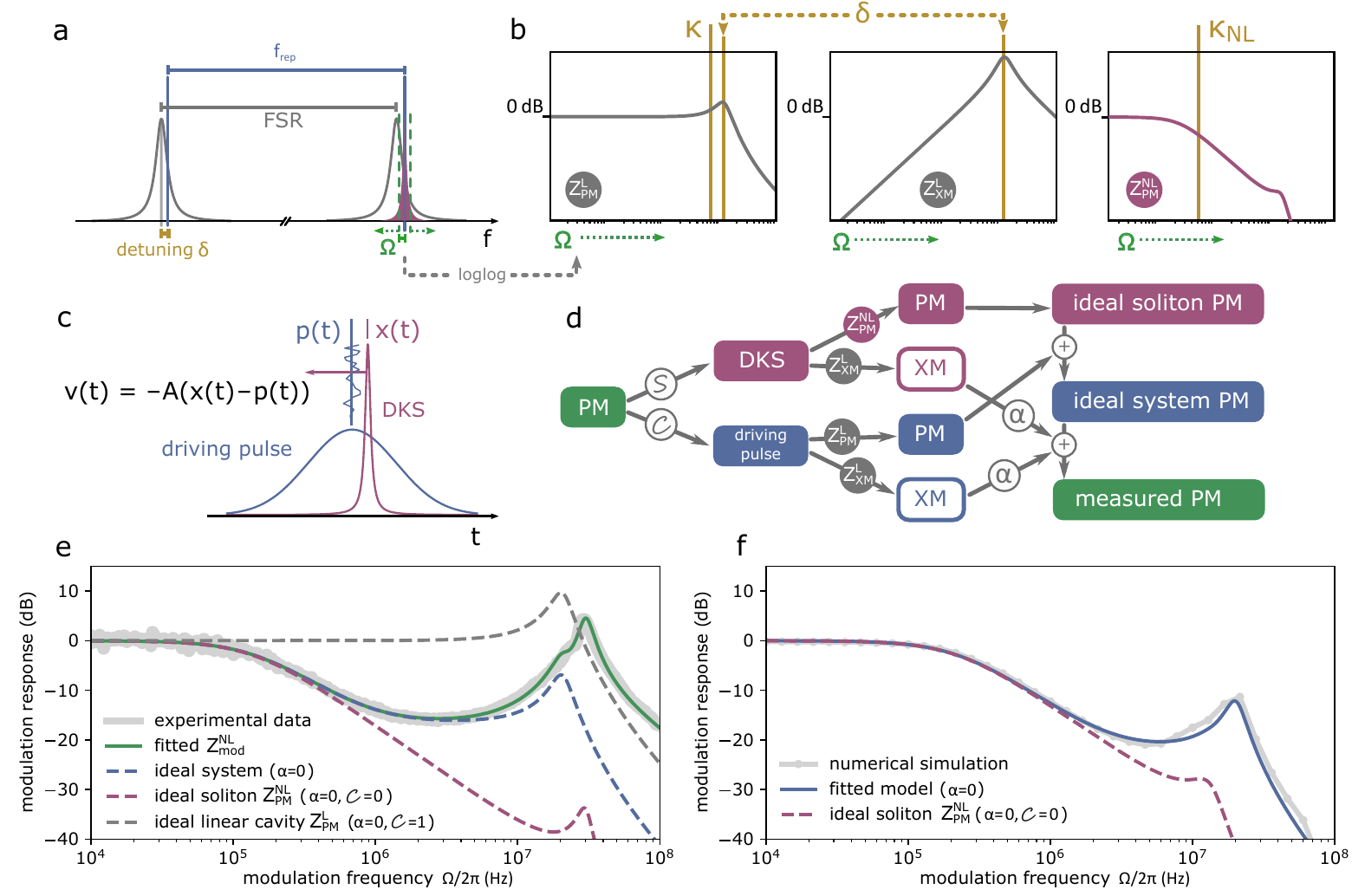}
\caption{\textbf{Nonlinear filter model. a} The linear filter is given by the Lorentzian lineshape (grey) of the resonator with linewidth $\kappa$ taking into account the detuning (gold) resulting in a more complicated lineshape shown in \textbf{b}. The nonlinear filter represents a much narrower Lorentzian (purple) with linewidth $\kappa_\mathrm{NL}$ which remains centered on the optical input frequencies for all detunings. 
\textbf{b} Shape of the different linear and nonlinear contributions. 
\textbf{c} Time domain picture of the soliton on top of the driving pulse and the resulting dynamic for the soliton. 
\textbf{d} Diagram showing the different modulation signal paths through the system and how they contribute. 
\textbf{e} Fit of the full model $Z_\mathrm{mod}^\mathrm{NL}$ (green) to the experimental data (solid grey, same data as green in Fig.~\ref{fig:fig2}d). Dashed lines represent the individual contributions as shown in \textbf{d}. Grey dashed is the pure linear response function.
\textbf{f} Phase modulation response function derived from numerical simulations (grey) and fitted model (blue). Dashed purple is again the ideal soliton PM response.
}
\label{fig:fig4}
\end{figure*}
\section{Analytical model}
To understand the observations and the physics of our system, we develop an analytical model starting out from the linear case and later extending it to the nonlinear case with DKS. 

For a linear optical cavity the standard quantum Langevin approach \cite{gardiner_input_1985} can be used to derive the timing jitter or microwave phase modulation transfer
\begin{equation}
    Z_\mathrm{PM}^\mathrm{L}(\Omega, \delta) = \frac{ \kappa^2 \left( 4 \Omega^2 + 8\delta^2 + \kappa^2 \right)  + 16 \delta^4}  {\left( 4 \Omega^2 + 4 \delta^2 + \kappa^2 \right)^2 - 64\Omega^2\delta^2}
    \label{eq:Z_PM_L}
\end{equation}
which is equivalent to the absolute optical phase modulation transfer of a single optical line.
For completeness we note that the non-zero group-velocity dispersion (GVD) of the cavity and a small potential mismatch between the cavity's FSR and the driving pulse repetition rate $f_\mathrm{rep}$ implies that the detuning $\delta$ between cavity modes and driving laser lines is sightly dependent on their distance from the central modes $\omega_{p}$ and $\omega_{0}$, respectively. However, in our model we neglect this fact.

The characteristic form of $Z_\mathrm{PM}^\mathrm{L}$ for $\delta = 1.7 \kappa$ is shown in Fig.~\ref{fig:fig4}b, left. It might seem counterintuitive that $Z_\mathrm{PM}^\mathrm{L}$ exceeds 0~dB for $\Omega\approx\delta$ but this can be understood by realizing that for a detuned, off-resonant line its modulation sideband can be resonant and gain relative power. 

In an \textit{ideal} experimental system timing modulation or phase modulation (PM) of the input pulse train will not translate into amplitude modulation (AM) in the intracaviy pulse. Indeed the optical frequency components of the input pulse train will ‘breath’ symmetrically around the pulse's center frequency such that a resulting intracavity amplitude modulation in a single sideband with relative mode index $+\mu$ (with regard to the central mode) will be compensated by the amplitude change in the sideband with mode index $-\mu$. However, in the case of inevitable imperfections such as a mismatch between the driving pulse repetition rate $f_\mathrm{rep}$ and the cavity's FSR (Fig.~\ref{fig:fig4}a), non-zero higher order dispersion and asymmetric driving spectra, even a pure timing modulation will result in some amplitude modulation. Moreover, AM-to-PM conversion caused by chirped driving pulses as well as AM-to-PM conversion in the photodetector and in the MW mixers can add an additional component to the measured signal. While it is difficult to quantitatively capture all these \textit{nonideal} effects, their functional shape is distinctively different from $Z_\mathrm{PM}^\mathrm{L}$ and described by the PM-to-AM transfer of a single optical line (again derived using the standard quantum Langevin approach)
\begin{equation}
        Z_\mathrm{XM}^\mathrm{L}(\Omega, \delta) = \frac{1024 \cdot D_1^2 \kappa^2 \Omega^2 \delta^2}{(4 \delta^2 + \kappa^2)^2 \cdot \left( ( 4 \Omega^2 + 4\delta^2 + \kappa^2)^2 - (8\Omega \delta)^2 \right)}
        \label{eq:Z_AM_L}
\end{equation}
where $D_1 = 2 \pi \cdot \mathrm{FSR}$. The characteristic form of $Z_\mathrm{XM}^\mathrm{L}$ is shown in Fig.~\ref{fig:fig4}b, middle, again for $\delta = 1.7 \kappa$.
Based on equations~\ref{eq:Z_PM_L} and \ref{eq:Z_AM_L}, we build a model that does not only include the \textit{ideal} timing fluctuation transfer characteristics represented by $Z_\mathrm{PM}^\mathrm{L}$ but in addition a \textit{nonideal} component with the functional form of $Z_\mathrm{XM}^\mathrm{L}$:

\begin{equation}
  Z_\mathrm{mod}^\mathrm{L}(\Omega, \delta) = Z_\mathrm{PM}^\mathrm{L}(\Omega, \delta) + \alpha Z_\mathrm{XM}^\mathrm{L}(\Omega, \delta)
  \label{eq:linmodel}
\end{equation}

where $\alpha$ is an empirical positive and unit-free parameter describing the effects resulting from experimental imperfections as discussed above.

Taking $\alpha$ and $\delta$ as the only free fitting parameters ($\kappa / 2\pi = 7.0$~MHz, $D_1/2\pi = 9.8$~GHz), we can model the measured trace $Z_\mathrm{meas}^\mathrm{L}$ (Fig.~\ref{fig:fig2}c, golden trace) very well for $\alpha = 3.2\cdot10^{-4}$ and $\delta = 1.7 \kappa$ (dashed-dotted trace). Based on this model, we can also plot the \textit{ideal} modulation response $Z_\mathrm{PM}^\mathrm{L}$ (dotted grey line in Fig.~\ref{fig:fig2}c). 

In order to extend the model to the nonlinear case, we first focus on the most striking feature in the nonlinear response, the suppression of modulations below the linear linewidth $\kappa$. This is surprising as, given the ultra-fast nature of the Kerr-nonlinear effects, the cavity decay time $1/\kappa$ can be assumed to be the slowest relevant time scale. On the other hand, it is known that DKS can exhibit slow dynamics resulting from weak soliton-soliton interaction \cite{jang_ultraweak_2013}, interaction with a chirped background \cite{jang_temporal_2015} or from interaction with a driving pulse \cite{obrzud_temporal_2017, hendry_spontaneous_2018, hendry_impact_2019}. 

Recent studies have shown that while in a stable equilibrium the DKS is bound to the driving pulse, it may drift with a certain relative velocity $v$ relative to the driving pulse if the system is perturbed \cite{hendry_spontaneous_2018, hendry_impact_2019} (Fig.~\ref{fig:fig4}c):
\begin{equation}
v = \frac{\mathrm{d}x}{\mathrm{d}t} = a(S, \delta) \frac{\mathrm{d}S}{\mathrm{d}x}+d
\label{eq:drift}
\end{equation}
where $t$ denotes time, $x(t)$ is the relative position of the soliton with regard to its bound equilibrium position on the driving pulse, $a$ is a coefficient that depends on the detuning $\delta$ and the local envelope of the driving field $S$, and $d$ is the drift coefficient due to the mismatch between the driving pulse repetition rate and the intrinsic repetition rate of the soliton state. The computation of $a$ can be performed numerically \cite{hendry_spontaneous_2018}.

Here we assume that the disturbances of the bound state due to the induced modulations or other perturbations are small such that the response of the system can be expanded and that it will be dominated by the linear part with coefficient $A = -\frac{\mathrm{d}v}{\mathrm{d}x}|_{x=0}$, where for a stable equilibrium $A > 0$.
The phase modulation introduced in our experiment is equivalent to a small timing modulation of the driving pulses, which we can describe as a harmonic perturbation of the driving pulse position $p(t) = \varepsilon \mathrm{e}^{i\Omega t}$, where $\Omega$ and $\varepsilon$ are the frequency and amplitude of the perturbation. The dynamics of the soliton are then described by the following differential equation
\begin{equation}
    \frac{\mathrm{d} x}{\mathrm{d}t} = -A (x(t) - \varepsilon \mathrm{e}^{i\Omega t}) \,
\end{equation}
which corresponds to a Lorentzian low-pass filter
\begin{equation}
    Z^\mathrm{NL}(\Omega) = \frac{A^2 }{\Omega^2 + A^2}
    \label{eq:znl}
\end{equation}
with a cut-off frequency of $\kappa_\mathrm{NL} = 2A$. This explains the observed nonlinear filter characteristics and shows that effectively, the nonlinear filtering can be interpreted as an equivalent additional dispersion-free, zero-detuned linear cavity filter with linewidth $\kappa_\mathrm{NL} < \kappa$. The cut-off frequency is determined by the parameter $A$ that depends on the laser detuning $\delta$, the repetition rate mismatch and the derivative of the driving pulse profile $\frac{dS}{dx}$ \cite{hendry_spontaneous_2018, hendry_impact_2019}.

One interpretation of this narrower filter bandwidth is, that effectively the nonlinear optical DKS dynamics can store information about the system's physical state much longer than the intrinsic cavity decay time $1/\kappa$ thereby enabling a filter bandwidth significantly below $\kappa$. 

The purple dashed lines in Figure~\ref{fig:fig2}d represent the best fits of the Lorentzian filter behavior $Z^\mathrm{NL}$ for low frequencies where linear cavity filtering effects of $Z_\mathrm{PM}^\mathrm{L}$ and $Z_\mathrm{AM}^\mathrm{L}$ are negligible. As expected from eq.~\ref{eq:drift}, $\kappa_\mathrm{NL}$ is detuning dependent (Fig.~\ref{fig:fig2}d), where a change in the detuning also impacts the repetition rate mismatch $\Delta$ and thereby the drift coefficient $d$. For a fixed pulse repetition rate $f_\mathrm{rep}$, the lowest value of $\kappa_\mathrm{NL}$, measured while changing the detuning, is more than 20 times below the cavity linewidth $\kappa$. This minimum corresponds to the green traces in Figure \ref{fig:fig2}d and f and in particular to the case where the envelope of the central portion of the spectrum is close to horizontal. 

In order to describe the full measured nonlinear trace also for higher modulation frequencies $\Omega$, we extend our model for the linear case (eq.~\ref{eq:linmodel}) to take into account the DKS with the effects described above as well as the driving pulses as summarized in Figure~\ref{fig:fig4}d.
First, we need to differentiate between the contribution of the residual driving pulses and the contribution of the DKS. Because of the large difference in intensity, the two parts experience different effective detunings $\delta_\mathcal{C}$ and $\delta_\mathcal{S}$ respectively. Here, our nomenclature adopts the convention established for CW driven DKS, where the different Kerr-shifts lead to distinctly detuned CW $\mathcal{C}$ and soliton $\mathcal{S}$ resonances \cite{guo_universal_2016}.

For the fraction $\mathcal{C}$ of the transmitted input light, we use the linear model developed above. For the soliton fraction $\mathcal{S}$ we combine the nonlinear model and the linear model by multiplying $Z_\mathrm{PM}^\mathrm{L}$ with $Z^\mathrm{NL}$ to obtain the full nonlinear PM response function $Z_\mathrm{PM}^\mathrm{NL}(\Omega, \delta_\mathcal{S})=Z^\mathrm{NL}(\Omega) Z_\mathrm{PM}^\mathrm{L}(\Omega, \delta_\mathcal{S})$ as shown in Fig.\ref{fig:fig4}b, right. As for the linear case (eq.~\ref{eq:Z_AM_L}), the response resulting from imperfections is given as $\alpha \cdot Z_\mathrm{XM}^\mathrm{L}$. Combining the linear with the nonlinear part the full model reads:
\begin{equation}
\begin{split}
  Z_\mathrm{mod}^\mathrm{NL}(\Omega, \delta_\mathcal{S}, \delta_\mathcal{C}) =  \mathcal{S}\cdot \left( Z^\mathrm{NL}_\mathrm{PM}(\Omega, \delta_\mathcal{S})
  + \alpha Z_\mathrm{XM}^\mathrm{L}(\Omega, \delta_\mathcal{S}) \right) \\ + \mathcal{C}\cdot \left(Z_\mathrm{PM}^\mathrm{L}(\Omega, \delta_\mathcal{C}) + \alpha Z_\mathrm{XM}^\mathrm{L}(\Omega, \delta_\mathcal{C}) \right)
  \label{eq:nonlinmodel}
\end{split}
\end{equation}
where $\mathcal{S}$ and $\mathcal{C}=1-\mathcal{S}$ are the relative contributions of the soliton and non-soliton fractions.
Figure~\ref{fig:fig4}e shows a fit of this nonlinear modulation transfer model (green) to the experimental trace (grey) where the following fit parameters were found: $\delta_\mathcal{C}/2\pi = 20.5$~MHz, $\delta_\mathcal{S}/2\pi = 30.0$~MHz, $\kappa_\mathrm{NL}/2\pi = 280$~kHz, $\alpha = 2.8\cdot10^{-5}$ and $\mathcal{S} = 0.978$. Despite the simple assumptions the model for $Z_\mathrm{mod}^\mathrm{NL}$ shows very good agreement with the experimental data, suggesting that it captures the essential underlying physics. 
The model therefore allows estimating the nonlinear filter characteristics for an ideal DKS based filter by making $ \alpha \rightarrow 0 $ and therefore rejecting effects related to imperfections as discussed above. The resulting trace is shown in Figure~\ref{fig:fig4}e in blue. Comparing this trace to the ideal linear filter characteristics ($\mathcal{C}=1$ and $\alpha=0$, dashed grey line) the strong modulation suppression by the DKS of several orders of magnitude is apparent in particular at low offset frequencies, where most oscillators are inherently noisy.
In addition, we can set $\mathcal{S}=1$ and $\mathcal{C}=0$ corresponding to a photodetection of only the soliton component (e.g. by perfectly suppressing the driving portion of the spectrum), resulting in even stronger timing modulation suppression (purple).

\section{Numerical simulations}
In order to further support our experimental and analytical results we perform numerical simulations that are based on the coupled mode model for synchronously driven microresonators \cite{chembo_modal_2010, hansson_numerical_2014, obrzud_temporal_2017}, with parameters close to those in the experiment.
The experiment, whose results are shown in Figure~\ref{fig:fig2}d, is repeated numerically by creating a DKS state and injecting an artificial timing modulation. Numerical demodulation leads to the modulation response shown in Fig.~\ref{fig:fig4}f. All characteristic features of the experiment are reproduced on a qualitative level demonstrating that also the numerical simulation captures the relevant physics. As the numerical simulation implements an idealized experiment there is good agreement of the simulated modulation response and the idealized model $Z_\mathrm{mod}^\mathrm{NL}$ with $\alpha=0$ (Fig.~\ref{fig:fig4}f, blue trace). The purple trace in Figure~\ref{fig:fig4}f shows $Z_\mathrm{mod}^\mathrm{NL}$ with $\alpha=0$ and $\mathcal{C}=0$.

\begin{figure}[h!bp]
\centering
\includegraphics[width=\linewidth]{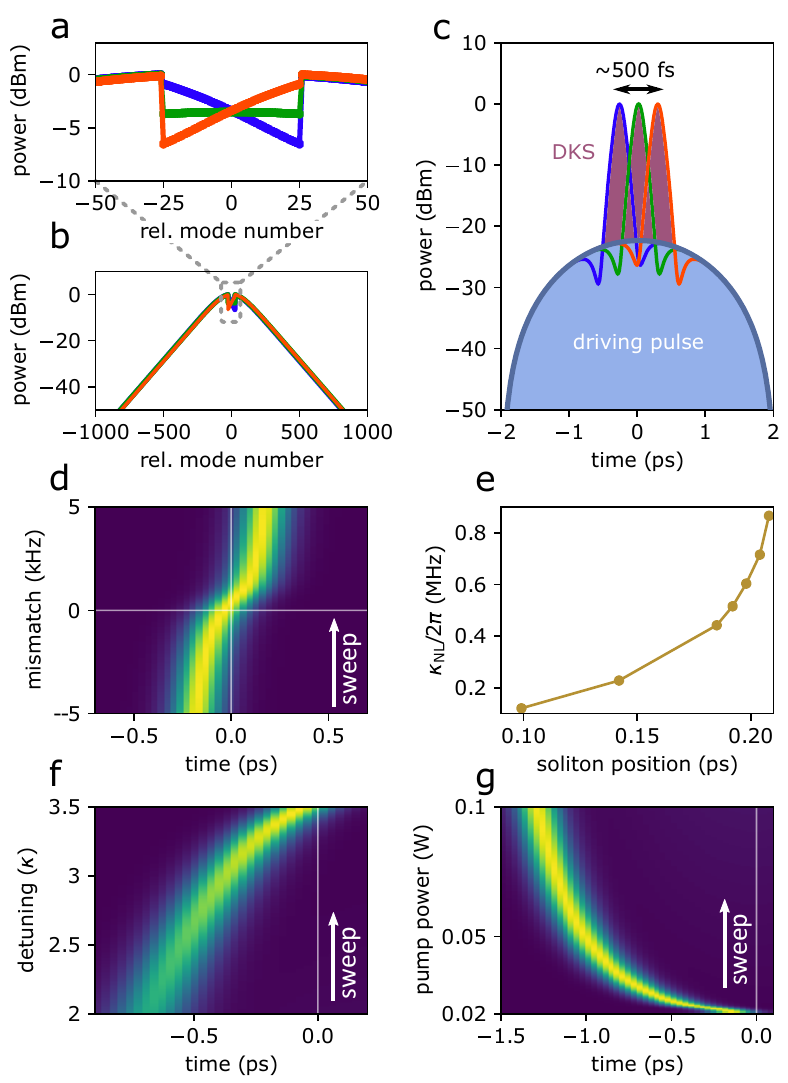}
\caption{\textbf{Soliton dynamics in numerical simulations. a} Zoom into the central part of the full simulated spectrum shown in \textbf{b}. \textbf{c} Soliton position on top of the driving pulse for the different cases shown in \textbf{a}. The position of the soliton in the middle of the driving pulse represents the symmetric spectrum (both in green). \textbf{d} Time domain intracavity field with respect to the driving pulse's temporal position.  As the repetition rate mismatch is numerically swept from --5~kHz to 5~kHz the soliton (high intensity, yellow) moves over the center of the pump pulse (white line at 0~ps). Other pump parameters: $\delta = $ 3.5 $\kappa$, $P_\mathrm{p} = $ 0.02~W. \textbf{e} Symmetric dependence of $\kappa_\mathrm{NL}$ on the soliton's temporal position relative to the driving pulse. \textbf{f} Sweep similar to \textbf{a} but with changing detuning ($\Delta = $~0~kHz, $P_\mathrm{p} = $~0.02~W). \textbf{g} As \textbf{d} and \textbf{f} but with changing pump power ($\Delta =$~0~kHz, $\delta = $~3.5~$\kappa$).
}
\label{fig:fig5}
\end{figure}

Numerically, creating a number of different soliton states, we find that the tilt of the spectral envelope as observed in Figure~2f is also reproduced in the simulations. Closer inspection shows that this tilt is directly related to the relative temporal position of the DKS with regard to the intracavity driving pulse. As shown in Figure~\ref{fig:fig5}, a perfectly flat spectral envelope (green trace in Fig.~\ref{fig:fig5}a and b) corresponds to the DKS being symmetrically centered on the underlying driving pulse (Fig.\ref{fig:fig5}c, driving pulse in blue, DKS in purple), tilted spectral envelopes are observed for off-centered DKS (red and blue traces), where the tilt increases for larger asymmetries. 
This, together with the insights from Fig.~\ref{fig:fig2}f and g, indicates a link between a more centered DKS position on the driving pulse and lower values of $\kappa_\mathrm{NL}$.

In order to better understand the effects of repetition rate mismatch, detuning as well as pump power on the DKS position and $\kappa_\mathrm{NL}$, we perform numerical simulations wherein each parameter is swept and studied separately free of thermally induced cross-coupling.

As shown in Figure~\ref{fig:fig5}d the DKS position can be tuned continuously via the repetition rate mismatch $\Delta$ across both sides of the driving pulse. Figure~\ref{fig:fig5}e confirms that more centered soliton positions are indeed linked to a lower $\kappa_\mathrm{NL}$.
Also, the detuning $\delta$ impacts the DKS position as shown in Figure~\ref{fig:fig5}f, where larger detunings (closer to the `edge' of the soliton step), result in more centered DKS. Finally, Figure~\ref{fig:fig5}g shows the dependence of the DKS position on the pump power, where a lower pump power leads to a more centered DKS position.

However, while perfectly centered DKS can be generated numerically, these states are found to be long-term stable only for low pump power levels and otherwise will assume a slightly off-center position (unless additional chirp is present in the driving pulses). This is in agreement with the experimentally observed switching behavior mentioned above and also consistent with previous analysis by Hendry et al. \cite{hendry_impact_2019} where for higher pump power, two symmetric stable off-center positions are identified. Which of those symmetric position is assumed is related to temporal symmetry breaking \cite{hendry_spontaneous_2018, xu_experimental_2014, copie_interplay_2019} and may be influenced via the repetition rate mismatch.

We conclude that a position of the DKS close to the center of the driving pulse results in lower values for $\kappa_\mathrm{NL}$ and that the DKS position itself depends on the combination of all pump parameters. For instance, in case of high pump power or small detuning a non-zero repetition rate mismatch may be used to obtain a more centered DKS position. An example for this are the results shown in Figure~\ref{fig:fig2}g, where the lowest $\kappa_\mathrm{NL}$ is obtained for a medium detuning for which the effective repetition rate mismatch $\Delta$ is optimal. In the same measurement, the increase of $\kappa_\mathrm{NL}$ for smaller (larger) detunings can be attributed to a thermally induced decreasing (increasing) mismatch $\Delta$, departing from the optimal condition.

\begin{figure}[ht]
\centering
\includegraphics[width=\linewidth]{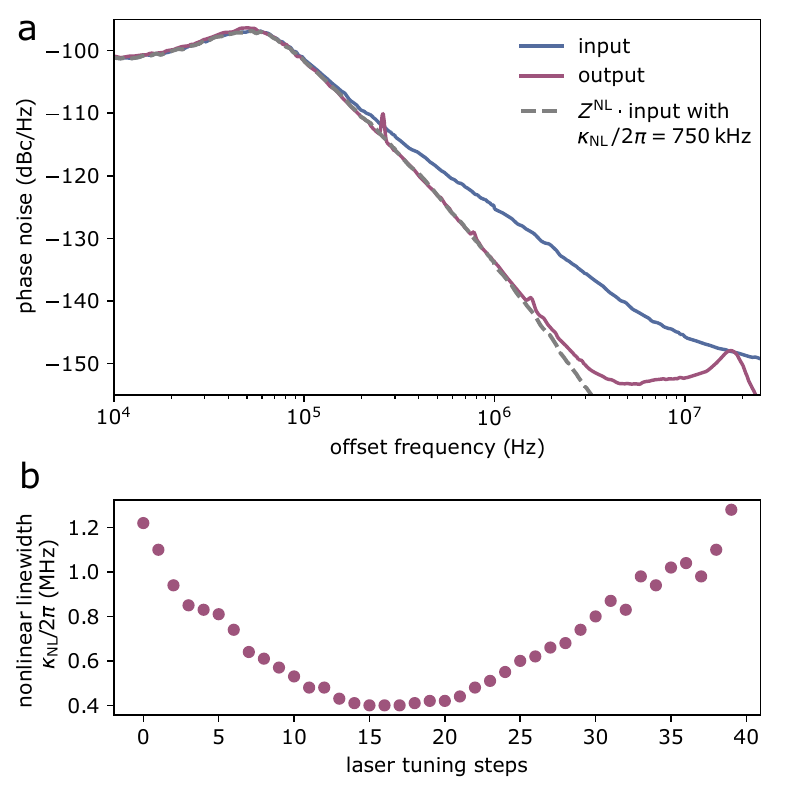}
\caption{\textbf{Absolute phase noise filtering. a} The intrinsic phase noise of a compact RF synthesizer is shown in blue as the phase noise of the input pulse train. The phase noise after nonlinear optical filtering is shown in purple. The grey, dashed trace is derived from the blue input trace by multiplying the data with $Z^\mathrm{NL}$ (eq.~\ref{eq:znl}) with $\kappa_\mathrm{NL}/2\pi = 750$~kHz. \textbf{b} Dependence of $\kappa_\mathrm{NL}$ on the detuning. As in Fig.~2g, the detuning is not quantified but simply represents different driving laser detuning steps as fixed by different laser lock points.}
\label{fig:fig3}
\end{figure} 
\section{Application}
It has been shown before that solitons driven by continuous wave pump lasers \cite{yi_soliton_2015, liang_high_2015, lucas_ultralow-noise_2019} can have very low phase noise at larger offset frequencies and, moreover, can be disciplined to a MW reference via weakly phase-modulated CW lasers without losing this purity \cite{weng_spectral_2019}. Here we show that using nonlinear filtering we can achieve corresponding phase noise reductions for a pulse train. 

In this measurement, a compact, low-cost microwave synthesizer is used to drive the electro-optic modulators that generate the pulse train. No additional timing modulation is injected and instead we measure directly the absolute phase noise of the pulse train before and after the cavity using a commercial, high-performance signal analyzer with cross-correlation function. We detect the entire output spectrum and do not compensate for any of the \textit{nonideal} contributions discussed above (therefore $\mathcal{S} < 1$ and $ \alpha > 0$). The measured phase noise after the microresonator and the DKS (purple trace in Fig.~\ref{fig:fig3}a) shows a clear suppression with a maximum of 15~dB at around 3~MHz compared to the input trace (blue), implying that the DKS acts as an effective clean-up oscillator. In order to quantify the nonlinear filtering behavior we fit $Z^\mathrm{NL}$ to the lower offset frequencies. For the data in Figure~\ref{fig:fig3}a we obtain a nonlinear linewidth $\kappa_\mathrm{NL}/2\pi$ of 750~kHz, 9$\times$ below the measured linear linewidth $\kappa/2\pi$ of the resonator.

Similarly to our previous measurement (Fig. \ref{fig:fig2}g) and our numerical simulations, we find again a detuning dependence of $\kappa_\mathrm{NL}$ with a minimum of approximately 400~kHz (Fig.\ref{fig:fig3}b.) 

We note that the decrease in phase noise suppression for higher offset frequencies (around 17~MHz in Fig. \ref{fig:fig3}a) is related to the detuning and the different \textit{nonideal} effects discussed above. The filter effect could be further increased by amplifying and detecting only light at wavelengths outside the input pulses' spectral coverage, which in equation \ref{eq:nonlinmodel} is equivalent to $\mathcal{S} \approx 1$ as transmitted residual input light is rejected.
\section{Conclusion}
In summary, we present a nonlinear optical filter based on dissipative Kerr soliton dynamics in a nonlinear optical microresonator, which reduces the timing jitter of an optical input pulse train. We show that the tunable bandwidth of the filter can be more than one order of magnitude lower than the cavity's intrinsic optical linewidth. In an example application we demonstrate that the soliton can act as a clean-up oscillator, leading to a phase noise, or equivalently timing jitter, reduction of an electro-optically generated pulse train, where the suppression can reach 15~dB for some offset frequencies.

These surprising effects are accurately described by an analytic model of the soliton dynamics and are fundamentally related to the solitons capability of storing information about the system's physical state longer than the cavity's light storage time. These insights can likely also explain previously made observations of sub-linewidth noise suppression, where a similar underlying mechanism might be in place.
Our findings are of immediate relevance to optical precision metrology, ultra-fast optical measurements, resonant supercontinuum generation in microresonators \cite{obrzud_temporal_2017, anderson_2019} as well as microwave photonics including ultra-narrowband optical filters for radar or signal processing.
The fact that DKS states can preserve information longer than the cavity's dissipation time scale may open new avenues for fundamental research at the interface of nonlinear optics and information theory.

\section*{Funding Information}
This work was supported by the Swiss National Science Foundation and Innosuisse via the Bridge Discovery Grant 20B2-1\_176563 as well as the Canton of Neuchâtel.

\section*{Acknowledgments}
We thank the Laboratoire Temps-Fréquence (LTF) at the University of Neuchâtel for support in performing the absolute phase noise measurements and Miro Erkintalo for helpful discussions. Moreover, we thank many members of the Open Source Software community for their great work as several different open source projects have been used in the course of the described work.

\nocite{*}
\selectlanguage{english}
\bibliography{NonlinearFiltering.bib}

\end{document}